\documentstyle[12pt]{article}
\begin{document}
\title{SYNCHRONIZATION AND CONTROL OF SPATIOTEMPORAL CHAOS USING
TIME-SERIES DATA FROM LOCAL REGIONS} 
\author{Nita Parekh, V. Ravi Kumar, and B.D. Kulkarni* \\
Chemical Engineering Division, \\ National Chemical Laboratory, \\ Pune -
411 008, INDIA}
\date{}
\baselineskip20pt
\parskip20pt
\maketitle
\vspace{2cm}
\begin{abstract}
In this paper we show that the analysis of the dynamics in localized
regions, i.e., {\it sub-systems} can be used to characterize the
chaotic dynamics and the synchronization ability of the spatiotemporal
systems.  Using noisy scalar time-series data for driving along with
simultaneous self-adaptation of the control parameter representative
control goals like suppressing spatiotemporal chaos and synchronization
of spatiotemporally chaotic dynamics have been discussed.
\end{abstract}

\noindent e-mail: ravi@ncl.ems.res.in
\pagebreak

\section{Introduction}

{\bf Most physical, chemical and biological systems are high-dimensional and
exhibit complex spatiotemporal patterns including spatiotemporal chaos
\cite{CH93}. The synchronization and control of the spatiotemporally chaotic
dynamics in these systems is currently being investigated and has been
reviewed in \cite{TS,SHOY,CD93,L95,LG94,M97,Lor97,Ar94,Peng,A94}.  In
this paper we study the synchronization and regulation of the
spatiotemporal systems using time-series data from local regions. This
approach may help in specifying the requirements of time-series data
from the spatial domain for control. Since the phase space is large for
spatiotemporal systems it may be worthwhile to first show
how the conventional diagnostics for low dimensional systems may be
appropriately utilized to study the synchronization behavior of higher
dimensional spatiotemporally chaotic systems. The feasibility of the
approach may be seen by studying the behavior of the sub-system
invariant properties such as the Lyapunov dimension and K-S entropy
\cite{CH94,Bauer,Arg94,Rev,KK,G89,MK89,W94} for increasing sub-system
size. Important and illustrative control goals, e.g., suppressing
spatiotemporal chaos by directing the system to desired stable fixed point or
low-period states (servo-control), and dynamical synchronization of the
spatiotemporally chaotic systems using localized {\it sub-system}
information have been addressed in this context.  The above aims have
been carried out for two prototype examples of coupled map lattices
(CMLs), {\it viz.}, the diffusively coupled logistic map (LCML) and the
diffusively coupled Henon map (HCML).}

The first CML studied is obtained by diffusively coupling $N$
logistic maps on a one-dimensional lattice \cite{KK} and is defined as
\begin{equation}
x_{n+1}(i) = (1-\epsilon)f(x_{n}(i)) + {\epsilon \over 2}
[f(x_{n}(i-1))+f(x_{n}(i+1))] ,
\end{equation}\noindent
where, $n$ is the discrete time; $i$ the lattice site, $i=1,2, \dots ,N$;
and $\epsilon$ the diffusive coupling coefficient. The nonlinear function
$f(x)$ is given by the quadratic form
\begin{equation}
f(x_{n}) \equiv x_{n+1} = 1 - \alpha x^2_{n} .  
\end{equation}\noindent
Equation (1) exhibits a wide variety of spatiotemporal patterns, {\it
viz.}, periodic, quasiperiodic, chaotic and complex frozen patterns
depending on the choice of parameters $\alpha$ and $\epsilon$
\cite{KK}. In Fig.~1(a) is shown the typical spatiotemporally chaotic
dynamics of the LCML (1) for periodic boundary conditions and {\it
random} initial conditions. This complex pattern arises due to
interactions between the diffusion and nonlinear mechanisms in the
LCML. The bifurcation parameter $\alpha = 1.9$ has been chosen such
that the local map (2) exhibits temporal chaos (Lyapunov exponent
$\lambda \sim 0.55$).  The coupling strength chosen was $\epsilon =
0.4$.

The second CML considered is the diffusively coupled Henon map lattice
in 1-D :
\begin{eqnarray}
x_{j,n+1}(i) & = & (1-\epsilon)f_j(x_{j,n}(i) + {\epsilon \over 2} 
[f_j(x_{j,n}(i-1) + f_j(x_{j,n}(i+1))] ,
\end{eqnarray}\noindent
where,
\begin{eqnarray}
x_{1,n+1} & \equiv & f_1(x_{1,n},x_{2,n}) = 1 - \alpha {x^{2}_{1,n}} +
x_{2,n} , \nonumber \\ x_{2,n+1} & \equiv & f_2(x_{1,n},x_{2,n}) = \beta
x_{1,n} ,
\end{eqnarray}\noindent
$j = 1, 2$; $i = 1,2,\dots N$. Again, the parameter values have been so
chosen that the local Henon map exhibits chaotic dynamics ($\alpha = 1.4$
and $\beta = 0.3$ for which the maximum Lyapunov exponent, $\lambda_{max}
\sim 0.42$) \cite{R89}. On assuming {\it identical} initial conditions for
$x_{j,0} (i)$, the HCML (3) exhibits spatially homogeneous but
temporally chaotic dynamics (as seen in Fig.~1(b) for $n < 100$). On
giving random perturbations to the central five lattice sites at $n =
100$, a changeover from spatially homogeneous to an inhomogeneous
spatiotemporal pattern is observed with the spread of perturbation to
the boundaries because of diffusive coupling [Fig.~1(b)]. The
following section discusses the analysis of spatiotemporally chaotic
dynamics in terms of the sub-system invariant measures. In section 3
the dynamical synchronization and control of spatiotemporal chaos in
these CMLs is discussed for representative goals.

\section{Analysis of Spatiotemporal Dynamics}

For a CML of size $N$, there are $mN$ Lyapunov exponents ($m$ being the
number of degrees of freedom in the corresponding single map, i.e.,
$m=1$ for logistic map and $m=2$ for Henon map) and their computation
can be taxing and practically infeasible for large $N$.  However,
if attention is restricted to a localized sub-system of size $n_s (<<
N)$, the calculation of the Lyapunov exponents is significantly reduced
to $mn_s$. The calculation of these sub-system exponents is similar to
those of the full system, that is, by time-averaging the growth rate of
linearized orthonormal vectors $\delta {\bf x}^l_n$, within the
sub-system, and is given by
\begin{eqnarray}
\lambda^{(s)}_l = \lim_{n \rightarrow \infty} \sup \hspace {2pt} \ln { {\mid
\delta {\bf x}^{l}_{n+1} \mid} \over {\mid \delta {\bf x}^{l}_0 \mid} },
\hspace {30pt} l = 1 \dots mn_s .
\end{eqnarray}\noindent
While calculating these exponents, the flow of information at the
sub-system boundary sites $k=1$ and $k=n_s$, may be treated as {\it
a})~noise effects, or, {\it b})~explicitly corrected by evaluating the
sub-system Lyapunov exponents only for $n_s-2$ sites (i.e., excluding
the boundary sites).  Our calculations of the sub-system Lyapunov
exponents, $\lambda^{(s)}_l$ for both treatments ({\it a,b}) were found
to be in quantitative agreement with open boundary conditions used for
the sub-system dynamics.

Now, from a knowledge of the spectrum of sub-system Lyapunov exponents,
$\lambda^{(s)}_i$, the effective sub-system Lyapunov dimension,
$d^{(s)}_L$, may be obtained and is defined as
\begin{equation}
d^{(s)}_L = j + {1 \over {\mid \lambda^{(s)}_{j+1} \mid }} \sum_{i=1}^j
{\lambda^{(s)}_i} ,
\end{equation}\noindent
on using the well-known Kaplan and Yorke (KY) conjecture \cite{KY79}. Here
$j$ is the largest integer for which the sum of the exponents,
$\lambda^{(s)}_1 + \dots + \lambda^{(s)}_j \geq 0$. If $\lambda_1 < 0$,
then $d^{(s)}_L = 0$ and if $j = mn_s$, then $d^{(s)}_L = mn_s$
\cite{R89}. The Lyapunov dimension gives the
effective dimensionality of the underlying attractor. The corresponding
intensive quantity, the sub-system dimension density, $\rho^{(s)}_d$,
may then be defined as
\begin{equation}
\rho^{(s)}_d = {d^{(s)}_L \over n_s}.
\end{equation}\noindent
It gives an estimate of the number of degrees of freedom required to
characterize the dynamical behavior of the full spatiotemporal system.  

Another important invariant measure is the Kolmogorov-Sinai (KS)
entropy and is defined as the sum of positive Lyapunov exponents
$\lambda_+$ \cite{P77}. It quantifies the mean information production
and growth of uncertainty in a system subjected to small perturbations
\cite{KY79}.  For regular predictable behavior,
the KS entropy is zero while for chaotic systems it takes a finite
positive value, and is infinite for continuous stochastic processes.
The sub-system KS-entropy, $h^{(s)}$, is defined as
\begin{equation}
h^{(s)} = \sum \lambda^{(s)}_+ .
\end{equation}\noindent
and the corresponding density function, the sub-system entropy density,
$\rho^{(s)}_h$, is given by
\begin{equation}
\rho^{(s)}_h = \sum \lambda^{(s)}_+ /n_s .
\end{equation}

The dependence of these invariant measures as a function of the
sub-system size, $n_s$ is discussed below.  In Fig.~2(a) is shown the
plot of the sub-system dimension, $d^{(s)}_L$, as a function of its
size $n_s$ (solid line corresponds to LCML and the dashed line to
HCML). The sub-system dimension $d^{(s)}_L$ is seen to linearly
increase with the sub-system size $n_s$ for both the CMLs. This
suggests that it may be possible to determine the effective
dimensionality of the whole system from sub-system analysis. Further,
the saturating behavior of the sub-system dimension density,
$\rho^{(s)}$ [Fig.~2(b)] helps in determining the critical sub-system
size, $n_{sc}$, required to predict the dimensionality of the full
system. Similar behavior was observed in the sub-system KS entropy,
$h^{(s)}$ and the entropy density, $\rho^{(s)}_h$, for increasing
sub-system size [Fig.~2(c) and 2(d)].  This implies that though the
entropy increases linearly with the sub-system size, the average rate
of information loss/gain levels off for $n_s > n_{sc}$. The above
relationships were also observed for logistic maps diffusively coupled
on a 2-dimensional square lattice of size $N \times N$ (results not
shown). These results indicate that it may be possible to analyze the
dynamical behavior of reaction-diffusion systems from an analysis of
relatively smaller sub-systems. This feature may prove to be
computationally very advantageous, especially in higher spatial
dimensions.

\section{Synchronization and Control of Spatiotemporal Chaos}

In this section, we discuss the synchronization and control of
spatiotemporally chaotic dynamics for different goals with the
following important factors considered, {\it viz.}, 1) a mechanism by
which a control parameter may be self-adapted so that synchronization
in the system and the desirable dynamics becomes possible; 2) allow for
restrictions in the availability of scalar time-series signals in the
spatial domain; and 3) negate the effects of noise in the time-series
data. From recent studies on the dynamical synchronization of
low-dimensional chaotic systems it is known that a given system (called
the {\it response}) can be made to follow the dynamics of another
system by driving the former with scalar time-series signals from the
latter \cite{FY83,PC90,HV92,DO94,CO93,WC94,H94}. The 
condition for the synchronization to occur is that the response system
should possess negative conditional Lyapunov exponents. Following the
results of Section 3, we would now like to see whether sub-system data
may suffice in assessing the synchronization ability of the
spatiotemporal system.

Before discussing the results, we present the methodology adopted
to synchronize the dynamics of spatiotemporal systems governed by
different attractors. For clarity we define the driving system by
\begin{equation}
{\bf x}_{n+1}(i) = {\bf F} [{\bf x}_{n}(i), {\bf x}_{n}(i\pm 1), \alpha, {\bf
\beta}] ,
\end{equation} \noindent
where ${\bf x}_n(i) = x_{j,n}(i)$, $j = 1, \dots, m$ ($m$ denotes the
number of degrees of freedom in the local map), and $i= 1,\dots , N$.
To incorporate the effects of noise arising due to measurement
errors, the sub-system driving signals are assumed to be given by
\begin{equation}
x'_{1,n} (k) = x_{1,n} (k) + \gamma \eta_n (k), \hspace {40pt} k = 1,
\dots, n_s ,
\end{equation} \noindent
where $\eta_n (k)$ denotes the random noise in the interval (-.5,.5) of
strength $\gamma$. The response system, written in a different notation
from eq. (10), is given by
\begin{eqnarray}
\hat {\bf x}_{n+1}(i) = {\bf F} [\hat {\bf x}_{n}(i), \hat {\bf x}_{n}(i \pm
1), {\bf x}'_{n}(k), \hat \alpha, {\bf \hat \beta}],
\end{eqnarray} \noindent
where $\hat {\bf x}_{n}(i)$ are the corresponding variables, $\hat
\alpha$ and $\hat \beta$ the response parameters, and ${\bf
x}'_{n}(k)$, the driving variables. To study the ability of the
response system to synchronize its dynamics with that of the driving
system (10), we analyzed the
conditional Lyapunov exponents for a localized sub-system ($n_s >
n_{sc}$). These exponents were calculated by monitoring the growth rate
of $(m-1)n_s$ sets of linearized orthonormal vectors obtained on
excluding the variables used for driving. For the HCML, the
calculations showed that the maximum sub-system conditional exponent is
negative on using $x_{1,n}(i)$ as the driving signals indicating
possible synchronization. On the other hand, if $x_{2,n}(i)$ were used
for driving, the maximum conditional exponent was found to be positive
and synchronization is not guaranteed. A synchronization study on HCMLs
with different initial conditions but same parameter values (i.e.,
$\hat \alpha = \alpha$, $\hat \beta = \beta $) did confirm the above
results. It may be also noted that in the case of LCML, the local map
being governed by a single variable (i.e., $m=1$) precludes the
observance of negative conditional Lyapunov exponents and
synchronization in their dynamics is difficult.

However, if driving is carried out on a response system with a
different setting of the control parameter, i.e., $\hat \alpha \neq
\alpha$, then synchronization of the response system dynamics cannot be
brought about by driving alone. In this situation, the control
parameter $\hat \alpha$ needs to be altered appropriately so that
synchronization becomes possible. Self-adaptive mechanisms for
parametric estimations have been studied in the context of temporal
chaotic systems \cite{HLSR,BD91,Q93,V94,P96}.  For spatiotemporal
systems, the self-adaptation of the control parameter may be carried
out as follows.  We begin by introducing a space-time dependence in the
response control parameter, i.e., $\hat \alpha_n (i)$.  Initially, the
same value of $\hat \alpha$ is assumed at all the lattice sites, but
different from that of the driving system, i.e., $\hat \alpha_0 (i) =
\hat \alpha \neq \alpha$. For the sub-system lattice sites where the
signals are available, the parametric corrections, $\Delta \hat
\alpha_{n+1}(k)$ may be dynamically evaluated as
\begin{equation}
\Delta \hat \alpha_{n+1} (k) = \Delta \hat \alpha_n (k) + \mu [\hat
x_{1,n} (k) - x'_{1,n} (k)], \hspace {40pt} k = 1, \dots, n_s ,
\end{equation}\noindent
where $\mu$ is the stiffness coefficient for adaptation and $\Delta
\hat \alpha_0 (i) = 0$. For the lattice sites outside the sub-system an
average adaptation, $\sum_k \Delta \hat \alpha_{n+1} (k) / n_s$, was
employed. The response parameter then self-adapts to the desired 
value $\alpha$ {\it via}
\begin{equation}
\hat \alpha_{n+1} {(i)} = \hat \alpha_0 (i) + \Delta \hat \alpha_{n+1}
(i), \hspace {30pt} i = 1, 2, \dots N .
\end{equation} \noindent
The linear functional form for adaptation considered in eq.~(13) is
only representative and other functional forms of adaptation, e.g.,
cubic, history-linear, sign, etc., \cite{HLSR,V94} may be attempted.
Further, the choice of $\mu$ may be rationalized by studying the
stability characteristics of the response and adapter dynamics. As long
as the combined system has negative eigenvalues synchronization should
be possible. A range of $\mu$ values can satisfy this requirement and
within this range the specific value of $\mu$ will determine the
rapidity with which synchronization occurs.

Using the above methodology, we discuss representative cases pertaining
to controlling spatiotemporal chaos. Our first aim was to suppress
chaos in a spatiotemporal system and direct it to a desired stable
fixed point state via self-adaptation of the control parameter along
with simultaneous driving. Noisy time-series signals (shown in
Fig.~3(a)) from a sub-system of size $n_s = 21$ localized in the
central region of the lattice of the driving system ($\alpha = 0.3$)
were used to drive the spatiotemporally chaotic dynamics of the
response system ($\hat \alpha_0 (i) = 1.9$). A rapid space-time
synchronization in the dynamics of the response and the driving system
is depicted by plotting the error signals $e_n(i) = \hat x_n(i) -
x'_n(i)$ in Fig.~3(b). The space-time convergent behavior of $\Delta
\hat \alpha_n (i)$ to a value of $-1.6$ (the initial difference in the
control parameter) by self-adaptation is shown in Fig.~3(c).  The
fluctuations in $\Delta \hat \alpha_n (i)$ is due to the presence of
noise in the driving signals which is constantly filtered by the
adapter eqs.~(13).  Thus, the simple form of self-adaptation given in
eq.~(13) can be effectively used even in the presence of reasonable
extents of noise to suppress chaos in the dynamics. The implementation
of the driving signals along with the adapter mechanism leads to a
faster convergence to the desired stable state. Further, because the
final state is a stable one, the system continues to operate in this
state even after the driving and self-adaptive mechanism have been
switched off. Similar results were also obtained for HCML using scalar
sub-system time-series signals (results not shown).  These results
suggest that it may be possible to suppress chaos in real experimental
situations by using scalar time-series signals from a local sub-system
with spatial self-adaptation of the control parameter. 

Next we considered controlling the spatiotemporally chaotic dynamics to
a temporally 2-period state. The sub-system time-series data from an
HCML exhibiting spatially homogeneous and temporally periodic
oscillations were used to drive the chaotic dynamics of the response
[shown in Fig.~4(a)]. On using the self-adaptive mechanism [eq.~(13)]
along with driving, the desired spatially homogeneous and temporally
periodic pattern is observed only within the sub-system [Fig.~4(b)].
Outside the sub-system, the dynamics is not phase synchronized, though
oscillating periodically in time.  On using driving signals from every
$5{th}$ lattice site, we were able to obtain complete synchronization
with phase locking in the response and the desired system dynamics
[Fig.~4(c)]. Thus, though local sub-system data is sufficient to
suppress chaos and direct the system to a stable periodic state, for
phase synchronization, time-series measurements from the full spatial
domain is required (which may be spaced out depending on the complexity
of the desired state). The space-time evolution of the parametric
correction is shown in Fig.~4(d).

The above studies were focused on directing the system to stable
states. Now we discuss the more difficult case of directing a
spatiotemporal system from one chaotic state to another. The results in
this study are presented for the HCML [eqs.~(3,4)] with sub-system
driving signals used to evaluate parametric corrections $\Delta \hat
\beta$ in
the alternate control parameter $\hat \beta$. This was carried out in a
procedure identical to evaluating $\Delta \hat \alpha$ and $\hat \alpha$
[eqs.~(13,14)] in Figs.~3,4.
In this case, synchronization was possible only within the sub-system
[Fig.5(a)], even though the response system control parameter had been
appropriately self-adapted [Fig. ~5(b)]. The asynchronous behavior
outside the sub-system is because of the sensitive dependence of the
chaotic dynamics to initial conditions and suggests that driving
signals from the entire spatial domain will be required for complete
synchronization in this case and was confirmed (results not shown).

Apart from estimating a control parameter, in many situations, it would
be desirable to accurately estimate other intrinsic system parameters.
Such a situation can arise when the other parameters of the response
system are not known {\it a-priori} \cite{Baake,V94,P96}. Here we show
that self-adaptation of two parameters is simultaneously possible on
using time-series signals only from a sub-system. In this study both
the response system parameters $\hat \alpha_0$ and $\hat \beta_0$ were
set differently from the driving system ($\alpha = 1.1, \beta = 0.3$,
$\hat \alpha_0 = 1.4, \hat \beta_0 = 0.28$).  The parametric corrections
$\Delta \hat \alpha$ and $\Delta \hat \beta$ were then simultaneously
estimated by using the following two sets of adapter equations within
the sub-system 
\begin{eqnarray}
\Delta \hat \alpha_{n+1} (k) & = & \Delta \hat \alpha_n (k) + \mu_1 [\hat
x_{1,n} (k) - x_{1,n} (k)] , \nonumber \\
\Delta \hat \beta_{n+1} (k) & = & \Delta \hat \beta_n (k) + \mu_2 [\hat
x_{2,n} (k) - x_{2,n} (k)] .
\end{eqnarray}\noindent
with $\mu_1$ and $\mu_2$ the stiffness coefficients for adaptation,
and $k = 1, \dots, n_s$. As before, average corrections, $\sum_k \Delta
\hat \alpha_{n+1} (k) / n_s$ and $\sum_k \Delta \hat \beta_{n+1} (k) /
n_s$ were implemented outside the sub-system. The simultaneous convergence
of $\Delta \hat \alpha_n \rightarrow - 0.3$ and $\Delta \hat \beta_n
\rightarrow 0.02$ in Figs.~6(a), 6(b) suggest that multiparameter
estimation may be possible in high-dimensional chaotic systems.
Although driving signals in both the variables in the sub- system were
necessary for accuracy, there exists a range of $\mu_1$ and $\mu_2$
values wherein multiparameter self-adaptation was successful.
Considerable potential exists in applying this technique in accurately
characterizing available mathematical models of an experimental system.
Using experimental time-series data from a sub-system, the parameter
values in the mathematical model (now the response system) can thus be
ascertained.

\section{Conclusion}

To summarize, the interesting scaling relationships that exist in the
sub-system invariant properties as a function of increasing sub-system
size have been used to study the synchronization properties of
high-dimensional spatiotemporal chaotic systems in a simpler fashion.
Our results show that suppressing spatiotemporal chaos and controlling
the system in desired stable fixed or low-period states is possible
using only sub-system data {\it via} self-tuning of a control parameter.
Simultaneous adaptation of more than one parameters using only
sub-system information is also possible. These results allow for
relaxation in the monitoring of time-series data from the spatial
domain for control purposes. On the other hand, the synchronization
studies with chaotic spatiotemporal dynamics suggest that
synchronization may be possible only in regions from which time-series
data is available.

{\bf Acknowledgments} We gratefully acknowledge the financial support of
the Department of Science and Technology, New Delhi, India, in carrying
out this work. One of the authors, NP would like to acknowledge CSIR,
India for financial support.

\pagebreak

\pagebreak

\begin{figure}

\caption{
\label{fig1}}
Spatiotemporal chaos in CMLs for lattice size $N = 100$ (every $10th$
iteration is plotted) : (a)
Spatiotemporal dynamics of LCML for $\alpha = 1.9$, $\epsilon = 0.4$.
Random initial conditions were assumed at $n = 0$. (b) Spatiotemporal
dynamics of HCML for $\alpha = 1.4$, $\beta = 0.3$, $\epsilon = 0.3$. At
$n = 0$ identical initial conditions were assumed. A finite random
perturbation given to the central five lattice sites, at $n = 100$
results in the complex spatiotemporal behavior because of coupling.
\end{figure}
\begin{figure}
\caption{
\label{fig2}}
Behavior of the sub-system invariants as a function of sub-system size
$n_s$ for LCML (solid line) and HCML (dashed line) : (a) Lyapunov
dimension, $d^{(s)}_L$; (b) dimension density, $\rho^{(s)}_d$; (c)
entropy, $h^{(s)}$; (d) normalized entropy, $\rho^{(s)}_h$. Parameters and
lattice size identical to fig.~1.
\end{figure}
\begin{figure}
\caption{
\label{fig3}}
Stabilization of the spatiotemporally chaotic dynamics with noise
reduction for the LCML. The response system ($\hat \alpha = 1.9$) was
assumed to be driven by noisy time-series signals, $x'_{1,n} (k)$, from a
sub-system ($n_s=21$) of the process ($\alpha = 0.3$).  (a) Measurement
noise, $\gamma \eta_n(k) = x'_{1,n} (k) - x_{1,n} (k)$, introduced in the
monitored sub-system process variables. (b) Space-time behavior of the
error signals, $e_1(i) = \hat x_{1,n} (i) - x_{1,n} (i)$, $i = 1,\dots,N$.
$e_1(i)$ is seen to fall to zero at all the lattice sites indicating
complete synchronization of the response dynamics with the process. (c)
Space-time plot of the adapter signals, $\Delta \hat \alpha$ implemented;
stiffness coefficient for adaptation, $\mu = 0.01$. Note that at $n = 0$,
the adapter $\Delta \hat \alpha = 0.0$ which then eventually assumes an
average value of $-1.6$ (the initial difference between $\alpha$ and $\hat
\alpha$). The adapter signals continuously filter the noise shown in (a)
to achieve the dynamical synchronization seen in (b).
\end{figure}
\begin{figure}
\caption{
\label{fig4}}
Controlling the spatiotemporally chaotic dynamics of a response system
($\hat \alpha = 1.4, \hat \beta = 0.3$) to temporally $2$-period state
($\alpha = 0.8, \beta = 0.3$). The results are shown for
HCML.  (a) Spatiotemporally chaotic dynamics of the response system
($\alpha = 1.4, \epsilon = 0.4$). (b) Spatiotemporal dynamics of the
response system on driving it with sub-system time-series signals
$x'_{1,n} (k)$, $k=41, \dots, 60$. Self-adaptive mechanism was
simultaneously implemented. (c) Oscillatory behavior with phase locking
in the spatial domain exhibited by the response system on driving it
with time-series signals $x_{1,n} (j)$, $j = 5, 10, \dots , 100$. (d)
Space-time convergence of $\Delta \hat \alpha$ to $- 0.6$ for $\mu = 0.001$.
\end{figure}
\begin{figure}
\caption{
\label{fig5}}
On using scalar time-series signals, $x_{1,n} (k)$, from a sub-system of
HCML exhibiting chaotic dynamics ($\alpha = 1.4, \beta = 0.3$) to drive
the response system ($\hat \alpha = 1.4$, $\hat \beta = 0.28$) operating
on a different chaotic attractor. (a) Till $n < 500$, the error between
the non-monitored process and response variables, $e_2 = x_{2,n} (i) -
\hat x_{2,n} (i)$, $i = 1, \dots, N$, is shown without driving or
adaptation. Synchronization is obtained only within the sub-system.
(b) Space-time behavior of the adapter signals
converging to $\Delta \hat \beta = 0.02$ (the initial difference between
$\beta$ and $\hat \beta$).
\end{figure}
\begin{figure}
\caption{
\label{fig6}}
Simultaneous estimation of both the parameters in HCML using only
sub-system time-series signals. The average parametric corrections,
$\Delta \hat \alpha_{av}$ and $\Delta \hat \beta_{av}$ implemented over
the entire spatial domain are shown. These values, respectively, converge
to $- 0.3 \pm 0.01$ and $0.02 \pm 0.001$ for $\mu_1 = 1.0$ and $\mu_2 = -
0.1$.
\end{figure}


\parskip10pt

\begin{thebibliography} {99}
\bibitem{CH93} M.C. Cross, and P.C. Hohenberg, Rev. Mod. Phys. {\bf
65}, 851 (1993).
\bibitem{TS} T. Shinbrot, Adv. in Phys. {\bf 44}, 73 (1995). 
\bibitem{SHOY} T. Shinbrot, C. Grebogi, E. Ott and J.A. Yorke, Nature {\bf
363}, 411 (1993).
\bibitem{CD93} G. Chen and X. Dong, Int. J. Bifur. Chaos {\bf 3}, 1363 (1993).
\bibitem{L95} J.F. Lindner, B.K. Meadows, W.L. Ditto, M.E. Inchiosa, and
A.R. Bulsara, Phys. Rev. Lett. {\bf 75}, 3 (1995).
\bibitem{LG94} Y.C. Lai and C. Grebogi, Phys. Rev. E {\bf 52}, 1894
(1994). 
\bibitem{M97} C.B. Muratov, Phys. Rev. E {\bf 55}, 1463 (1997).
\bibitem{Lor97} M.N. Lorenzo, I.P. Marino, V. Perez-Munuzuri, M.A.
Matias, and V. Perez-Villar, Phys. Rev. E {\bf 54}, R3094 (1997).
\bibitem{Ar94} I. Aranson, H. Levine, and L. Tsimring, Phys. Rev.
Lett. {\bf 72}, 2561 (1994).
\bibitem{Peng} J.H. Peng, E.J. Ding, M. Ding, and W. Yang, Phys.
Rev. Lett. {\bf 76}, 904 (1996).
\bibitem{A94} D. Auerbach, Phys. Rev. Lett. {\bf 72}, 1184 (1994).
\bibitem{CH94} M.C. Cross and P.C. Hohenberg, Science {\bf 263}, 1569 (1994).
\bibitem{Bauer} M. Bauer, H. Heng, and W. Martienssen, Phys. Rev.
Lett. {\bf 71}, 521 (1993).
\bibitem{Arg94} J. Argyris, G. Faust, M. Haase, {\it An Exploration of Chaos}
Elsevier Science B.V., Amsterdam (1994).
\bibitem{Rev} H.D.I. Abarbanel, R. Brown, J.J. Sidorowich, and L. Tsimring,
Rev. Mod. Phys. {\bf 65}, 1331 (1993).
\bibitem{KK} K. Kaneko, Prog. Theor. Phys. {\bf 72}, 480 (1984);
Physica D {\bf 23}, 436 (1986); Physica D {\bf 34}, 1 (1989); Chaos,
{\bf 2}, No. 3 (1992) (special issue on coupled map lattices) and
references therein.
\bibitem{G89} P. Grassberger, Phys. Scr. {\bf 40}, 346 (1989).
\bibitem{MK89} G. Mayer-Kress and K. Kaneko, J. Stat. Phys. {\bf
54}, 1489 (1989).
\bibitem{W94} J. Warncke, M. Bauer, and W. Martienssen, Europhys. Lett.
{\bf 25}, 323 (1994). 
\bibitem{R89} S.N. Rasband, {\it Chaotic Dynamics of Nonlinear Systems}
Wiley-Interscience (1989).
\bibitem{KY79} J.L. Kaplan, and J.A. Yorke, {\it Lecture Notes in
Mathematics} {\bf 730}, 204 (1979).
\bibitem{P77} Y.B. Pesin, Russ. Math. Sur. {\bf 32}, 55 (1977).
\bibitem{FY83} H. Fujisaka, and T. Yamada, Prog. Theor. Phys. {\bf
69}, 32 (1983).
\bibitem{PC90} L.M. Pecora, and T.L. Carroll, Phys. Rev. Lett. {\bf 64},
821 (1990); {\it Phys. Rev. A} {\bf 44} 2374 (1991).
\bibitem{HV92} R. He and P.G. Vaidya, Phys. Rev. A {\bf 46}, 7387 (1992).
\bibitem{DO94} M. Ding, and E. Ott, Phys. Rev. E {\bf 49}, R945 (1994).
\bibitem{CO93} K.M. Cuomo, and A.V. Oppenheim, Phys. Rev. Lett. {\bf
71}, 65 (1993).
\bibitem{WC94} C.W. Wu, and L.O. Chua, Int. J. Bifur. Chaos {\bf 4},
979 (1994).
\bibitem{H94} J.F. Heagy, T.L. Carroll, and L.M. Pecora, Phys. Rev.
E. {\bf 50}, 1874 (1994).
\bibitem{HLSR} B.A. Huberman, and E. Lumer, IEEE Trans. Circuits
Syst. {\bf 37}, 547 (1990); S. Sinha, and R. Ramaswamy, Physica D {\bf
43}, 118 (1990). 
\bibitem{BD91} V. Ravi Kumar, B.D. Kulkarni, and P.B. Deshpande, Proc.
R. Soc.  London Ser. A {\bf 433}, 711 (1991); J.K. Bandyopadhyay, V.
Ravi Kumar, B.D. Kulkarni, and P. Bhattacharya, Chem. Eng. Sci. {\bf
48}, 3545 (1993). 
\bibitem{Q93} H.K. Qammer, F. Mossayebi, and L. Murphy, Phys.  Lett.
A {\bf 178}, 279 (1993).
\bibitem{V94} D. Vassiliadis, Physica D {\bf 71}, 319 (1994).
\bibitem{P96} U. Parlitz, Phys. Rev. Lett. {\bf 76}, 1232 (1996); Int. J.
Bifur. Chaos, {\bf 6}, 581 (1996).
\bibitem{Baake} E. Baake, M. Baake, H.G. Bock, and K.M. Briggs, Phys.
Rev. A {\bf 45}, 5524 (1992).
\end{thebibliography}
\end{document}